\documentstyle[prb,multicol,aps,epsf]{revtex}
\begin{document}
\draft

\title{Metallic Nanosphere in a Magnetic Field:
an Exact Solution} \author{D.N. Aristov}
\address{ Petersburg Nuclear Physics Institute,
Gatchina, St. Petersburg 188350, Russia}\date{\today} \maketitle

\begin{abstract}
We consider the electron gas moving on the surface of a sphere in a
uniform magnetic field. An exact solution of the problem is found in
terms of oblate spheroidal functions, depending on the parameter $p=
\Phi/\Phi_0$, the number of flux quanta piercing the sphere.  The
regimes of weak and strong fields are discussed, the Green's functions
are found for both limiting cases in the closed form.  In the weak
fields the magnetic susceptibility reveals a set of jumps at
half-integer $p$.  The strong field regime is characterized by the
formation of Landau levels and localization of the electron states near
the poles of the sphere defined by a direction of the field.  The
effects of coherence within the sphere are lost when its radius exceeds
the mean-free path.
\end{abstract}


\begin{multicols}{2} \narrowtext

The electronic properties of cylindrical and spherical carbon
macromolecules \cite{fulle} have attracted much of theoretical interest
last years. A large part of these studies is devoted to the
 band structure calculations \cite{band} and to the effects of topology
for the transport and mechanical properties of these nanostructures.
The main interest to the topological aspects is connected with the
carbon nanotubes, where one can investigate the effective models, based
on the band calculations and incorporating the particular geometry of
the object. \cite{currents}

At the same time, the problem of the topology appears to
be a general one and may be studied independently from the physics of
carbon materials. The recent advances in technology \cite{SiO} let one
think of a wider class of the spherical nanostructures, e.g. the
spheres coated by metal films,\cite{coated} whose properties may differ
from those of planar objects.

In this paper we study the gas of electrons moving within a thin
spherical layer in an applied magnetic field. We find the exact
solution of this problem in terms of oblate spheroidal functions.
This physical application of the theory of spheroidal functions is
demonstrated apparently for the first time. \cite{Flammer,Me-Sch,KPSS}
We show the jumps in the susceptibility of the system in ``weak'' fields
and the localization of the electronic states in ``strong'' fields.
The last effect could be experimentally investigated for the
hemispherical tips of nanotubes. \cite{STM} At the intermediate
fields the sophisticated structure of the functions makes
the analytical treatment impossible and the numerical methods should be
used for the analysis of observable quantities.

\section{a setting-up of the problem}

Let us consider the electrons moving on a surface of the sphere of
radius $r_0$.  In the presence of the uniform magnetic field ${\bf B}$
we choose the gauge of the vector potential as a vector product ${\bf
A}=\frac12 ({\bf B}\times{\bf r})$.  The Hamiltonian of the system is
given by

     \begin{eqnarray}
     {\cal H} &=& \frac1{2m_e}
     (-i\nabla + e {\bf A})^2  + U(r),
     \end{eqnarray}

\noindent
where $m_e$ is the (effective) mass of an electron, $-e$ its charge;
we have set $\hbar=c=1$ and omitted the trivial term connected with the
spin of the particle.  We assume that the total number of particles $N$
(with one projection of spin) is fixed and defines the value of
the chemical potential $\mu$ and the areal
density $\nu = N/(4\pi r_0^2)$.  The confining potential $U(r) = 0$ at
$r_0< r <r_0 + \delta r$ and $U(r) \to \infty$ otherwise.  We focus our
attention on the limit $\delta r \ll r_0$, when the variables of the
problem are separated.  The radial component $R(r)$ of the wave
function is a solution of the Shr\"odinger equation with the quantum
well potential.  Henceforth we ignore the radial component and put
$r=r_0$ in the remaining angular part of the Hamiltonian ${\cal
H}_{\Omega}$ ; it can be done if $\mu$ lies below the first excited
level of $R(r)$, which in turn means $\delta r \lesssim \nu ^{-1/2}$.
We choose the direction of the field $\bf B$ along the $\widehat{\bf
z}-$axis and as a north pole of the sphere ($\theta =0$) and look for
the eigenfunctions to the Shr\"odinger equation ${\cal H}_{\Omega} \Psi
= E\Psi$ in the form $\Psi(\theta,\phi) = S(\theta) e^{im\phi} $.
Defining the dimensionless energy $\varepsilon = 2m_er_0^2 E $ and
setting $\eta=\cos\theta$, we write

     \begin{equation}
     \frac{\partial}{\partial \eta}
     (1-\eta^2) \frac{\partial S}{\partial \eta}
     + \left[ \varepsilon -2mp - \frac{m^2}{1-\eta^2}
     -  p ^2 (1-\eta^2)
     \right] S =0 .
     \label{sphdifeq}
     \end{equation}

\noindent
We introduced here the important dimensionless parameter
     \begin{equation}
      p={ e Br_0^2}/2 = {\pi Br_0^2}/{\Phi_0}
     ={ r_0^2}/({2 l_\ast^2}) =  m_e r_0^2 \omega_c/2,
     \label{def-p}
     \end{equation}
with the magnetic flux quantum $\Phi_0= 2\pi/e=2\cdot 10^ {-15}$
T$\cdot$m$ ^2$, the magnetic length $l_\ast = (eB)^{-1/2}$, the
cyclotron frequency $\omega_c=eB/m_e$. Note that for a sphere of radius
$r_0=10\,$nm one has $p=1$ at the field $B\simeq 6\,$T.

The equation (\ref{sphdifeq}) is known as the spheroidal differential
equation and was extensively studied previously.
\cite{Flammer,Me-Sch,KPSS} The solutions to it are given by the oblate
(angular) spheroidal functions $S_{lm}( p ,\eta)$ with the
corresponding eigenvalues $\varepsilon_{lm}(p)$.

It is known that spheroidal functions belong
to the simplest class of special functions which are not essentially
hypergeometric ones. For the spheroidal functions  there are no
recurrence relations, generating function representations etc., which
are characteristic for the classical special functions. The spectrum
$\varepsilon_{lm}(p)$ is found as the eigenvalues of the infinite
matrices. For different sets of functions orthogonal on the interval
$(-1,1)$, these matrices are reduced to the tridiagonal form and could
be further analyzed within the chain fraction formalism or by explicit
(numerical) diagonalization. \cite{Flammer,Me-Sch,KPSS}
Perhaps, the mostly known quantum-mechanical application of the
spheroidal functions is the problem of an electron in two-center
Coulomb potential (H$_2^+$ molecule).

\section{the weak field, susceptibility}

Let us first concentrate on the case of the weak field. In the
absence of the field, the spectrum is that of a free rotator model and
the solutions to (\ref{sphdifeq}) are the associated Legendre
polynomials \cite{fnote}:

        \begin{eqnarray}
        \varepsilon_{lm}(0 ) &=& l(l+1), \nonumber \\
        S_{lm}(0 ,\eta) &=& \sqrt{\frac{2l+1}{4\pi r_0^2}
        \frac{(l-|m|)!}{(l+|m|)!}} \,P_l^m(\eta).
        \label{zeroB}
        \end{eqnarray}
According to (\ref{zeroB}) the wave functions are the spherical
harmonics $\Psi_{lm}(\theta,\phi)= r_0^{-1} Y_{lm}(\theta,\phi) $ and
are normalized on the surface of the sphere, $r_0^2 \int
|\Psi|^2\sin\theta d\theta d\phi =1 $. This normalization facilitates
the comparison of our results with the case of planar geometry.

For $p\neq0$ one develops the perturbation theory around the
initial wave functions (\ref{zeroB}) as long as $p^2 \leq 4l $.
\cite{Flammer,Me-Sch,KPSS}
The energies and wave functions in this case are given by
series in $p^2$ :

     \begin{eqnarray}
     \varepsilon_{lm}(p) = && l(l+1) + 2 p m + \frac{p^2}2
     \left[1+\frac{m^2}{l^2}\right] + O\left[\frac{p^2}l\right],
     \label{ene-weak}        \\
     S_{lm}(p,\eta) && \propto P_l^m(\eta) +
     P_{l\pm2}^m(\eta) \, O\left[{p^2}/l\right]
     \label{wf-weak}
     \end{eqnarray}

To clarify the criterion $p^2/l \lesssim 1$ we consider the case when
the field ceases to be small for electrons near the Fermi level $l
\simeq r_0 \sqrt{4\pi\nu}$, mostly contributing to the physical
properties.  This case corresponds to the following relation

     \[
     Br_0^3 \sim \left[ {\hbar c}/{e^2}\right]^2
      e^2 \sqrt{\nu} \sim 137^2 \omega_{pl}.
     \]

\noindent
Hence the field cannot be treated as a perturbation,
if the energy of the magnetic field in the volume of the sphere is
$10^4$ times larger than the characteristic plasma frequency.
For densities $\nu\sim 10^{14}\,$cm$^{-2}$ and $r_0=10\,$nm it
corresponds to the fields greater than 40 T.

For the weak field regime it is interesting to observe the following
property of the spectrum (\ref{ene-weak}).
For simplicity we consider the situation when the $L$th unperturbed
level is completely filled and $(L+1)$th level is empty.
Linearizing the spectrum we have $\varepsilon_{lm} \simeq 2L
(l - L-1/2 + pm/L)$. It is clear that at $p>1/2$ the state
$(l=L+1,m=-L-1)$ is energetically more favorable than the state
$(l=L,m=L)$, with the resulting change in the occupation of the levels.

\begin{figure}
\centerline{\epsfxsize=8cm
\epsfbox{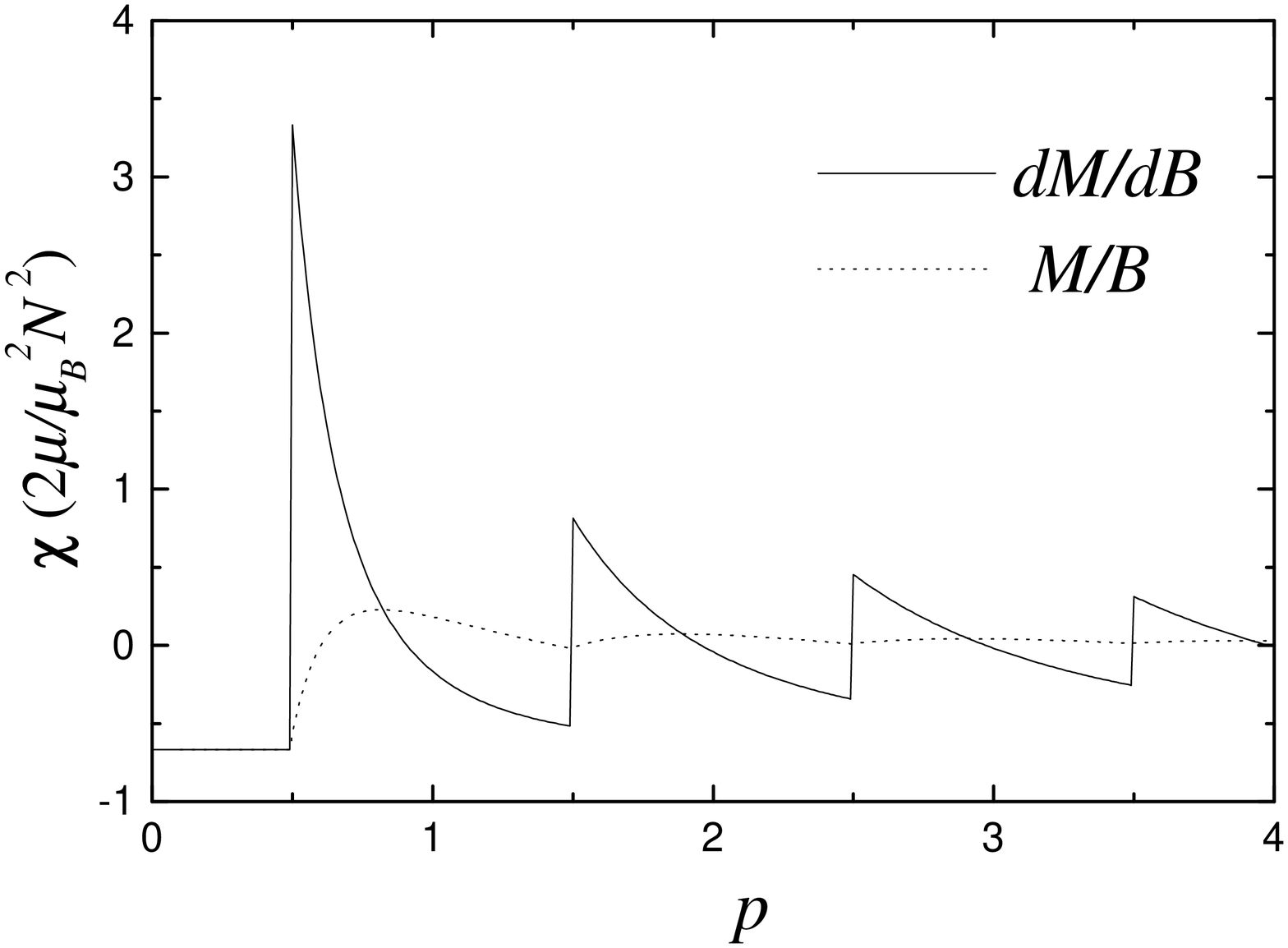}}
\caption{ The dependence of the static susceptibility $\chi$ on the
number $p$ of magnetic flux quanta piercing the sphere.
\label{fig:susc}  } \end{figure}

This phenomenon is accompanied by the jumps in the static
susceptibility. Consider the free energy $F = -T
\sum_{lm} \ln(e^{(\mu - E_{lm})/T}+1 )$. Then, apart from the
Pauli spin contribution, the magnetic (differential) susceptibility
$\chi =\partial M/\partial B = -\partial^2 F/\partial B^2$ is
given by

        \[
        \chi = -\frac{\mu_B^2m_er_0^2}2 \sum_{lm}\left[
        \frac{\partial^2 \varepsilon_{lm}} {\partial p^2}
        n_F(\varepsilon_{lm})
        +\left[\frac{\partial \varepsilon_{lm}} {\partial p}
        \right]^2 n'_F(\varepsilon_{lm})\right]
        \]

\noindent
with the Bohr magneton $\mu_B=e/2m_e$, the  Fermi function
$n_F(\varepsilon)$ and $n'_F(\varepsilon)$ its derivative.  The first
term here gives the diamagnetic contribution $\chi^d \simeq
-2/3 \mu_B^2 m_er_0^2N$ and the second term is the paramagnetic one.
 At low temperatures $\omega_c \lesssim T\ll\mu$ we let
$n'_F(E_{lm})\simeq -\delta(\varepsilon_{lm}-\mu)$ and change $\sum_m
\simeq \int_{-l}^l dm$.  Performing the integration we get

        \begin{equation}
        \chi =  N^2 (\mu_B^2/2\mu) \left[ -2/3 +
        p^{-3} \sum\nolimits_l (l-L-1/2)^2  \right]
        \label{susc0}
        \end{equation}

\noindent
where the summation is restricted by $|l-L-1/2| <p $. From (\ref{susc0})
we see that $\chi$ exhibits the jumps at $p=1/2,3/2,5/2\ldots$ while
$\chi \to 0$ in the formal limit  $p\to \infty$. This behavior is shown
in the Fig.\ \ref{fig:susc}, together with the quantity $M/B$. At the
other fillings, i.e.  at the other values of $\mu$ the jumps of $\chi$
take place at other values of $p$ and the qualitative picture remains
true.
The amplitude of the jumps is $N$ times larger than the
Pauli spin susceptibility $\sim N \mu_B^2/\mu$; this coherent effect
vanishes if the coherence on the sphere is lost due to
the finite quasiparticle lifetime (see below).

\section{the weak field, Green's function}

Let us discuss the properties of the electron Green's function. For the
two points on the sphere ${\bf r}\leftrightarrow (\theta,0)$ and
${\bf r}'\leftrightarrow (\theta',\phi)$ we define

     \begin{equation}
     G({\bf r},{\bf r}',\omega) =
     \sum_{lm}\frac{ \Psi_{lm}^\ast(\theta',\phi) \Psi_{lm}(\theta,0)
     }{\omega +\mu -E_{lm}} ,
     \label{defG}
     \end{equation}
with $E_{lm}= (2m_er_0^2)^{-1} \varepsilon_{lm}$.

In the absence of the magnetic field $\Psi_{lm}(\theta,\phi) = r_0^{-1}
Y_{lm}(\theta,\phi)$ and one can find (see below) an exact
representation of $G$ through the Legendre function

     \begin{equation}
     G(\omega)_{B=0} \equiv
     G^{0}(\omega) =
     -\frac{m_e}{2\cos \pi a} P_{-1/2+a}(-\cos\Omega),
     \label{G0}
     \end{equation}
where we introduced $a = \sqrt{2m_er_0^2(\mu+\omega)+1/4}$ and the
distance $\Omega$ on the sphere :

     \[
     \cos\Omega =
     \cos\theta \cos\theta' +
     \sin\theta \sin\theta' \cos\phi
     \]

\noindent
We immediately see from (\ref{G0}) the logarithmic singularity in the
one-point correlator ($\Omega=0$) as it should be.  It is instructive
to find out how the usual expressions for the planar geometry
are recovered for the large radius of the sphere, $r_0\to \infty$; in
this case $a\propto r_0$, while $\Omega \to |{\bf r}-{\bf
r}'|/r_0\equiv r/r_0$.  In the limit $a\gg1$ and for $a\sin\Omega
\gtrsim 1$ one has  in the main order of $a^{-1}$ : \cite{Ba-Er}

     \begin{equation}
     G^{0}(\omega) \simeq
     - \frac{m_e}{\sqrt{2\pi a\sin\Omega}}
     \frac{\cos(a\pi-a\Omega-\pi/4)}{\cos \pi a}  ,
     \label{G0-asymp}
     \end{equation}

\noindent
The existence of two oscillating exponents $\exp\pm i(a\pi- a\Omega-
\pi/4)$ corresponds to quantum coherence of two waves. One propagates
along the shortest way between two points and another wave goes along
the longest way, turning round the sphere. In the theory of metals
one considers $|\omega|\ll \mu= k_F^2/(2m_e)$,
while the finite quasiparticle lifetime can be modelled by
ascribing the imaginary part to $\omega$.  We write in this sense
$\sqrt{2m_e\omega}=k_F+i/l_{mfp}$ with the mean free path $l_{mfp}$. When
$r_0\gtrsim l_{mfp}$, we see that the coherence breaks and the only
surviving exponent in (\ref{G0-asymp}) has the form

     \begin{equation}
     G^{0}_{damped}(\omega) \simeq
     -\frac{m_e }{ \sqrt{2\pi k_Fr}}
     \exp\left[ik_Fr+i\frac\pi4-\frac r{l_{mfp}}\right],
     \label{G0-damped}
     \end{equation}
which expression coincides with the usual findings. \cite{ar}

It is possible to obtain the closed form of the Green's
function in the weak field regime. We sketch the corresponding
derivation below.

First we neglect the $p^2/l$ terms in the wave function (\ref{wf-weak})
and write
     $
     Y_{lm}^\ast(\theta',\phi) Y_{lm}(\theta,0)
     = (4\pi)^{-1} (2l+1) e^{-im\phi}
     P_l^m(\cos\theta) P_l^{-m}(\cos\theta').
     $
Using (\ref{ene-weak}) and dropping the terms containing $p^2/a\sim
p^2/l$, we have

     \[
     \frac{2l+1}{\omega - \varepsilon_{lm}} \simeq
     \left[a + \frac{pm}a - l-\frac12\right]^{-1} -
     \left[a + \frac{pm}a + l+\frac12\right]^{-1}
     \]

\noindent
Now we represent the appearing fraction as the Taylor series
     $ (a + pm/a - z)^{-1} =
     \exp\left( \frac{pm}{a}\frac\partial{\partial a}\right)
     (a - z)^{-1} $
and substitute $m$ by $i\frac\partial{\partial \phi}$ in the last
exponent. After that we can sum over $m$ in (\ref{defG}) \cite{Ba-Er}
and represent the remaining sum over $l$ as the contour integral, to
obtain

     \begin{eqnarray}
     G(\omega) &=&
     \exp\left[i\frac{p}{a}
     \frac{\partial^2}{\partial a\partial \phi }\right]
     \oint \frac{d\nu}{2\pi i} \frac{m_e}{2 \sin \pi\nu}
     \frac{P_{\nu}(-\cos\Omega)}{a-\nu-1/2}
     \nonumber \\ &=&
     \exp\left[i\frac{p}{a}
     \frac{\partial^2}{\partial a\partial \phi }\right]
     G^{0}(\omega),
     \label{G-weak}
     \end{eqnarray}

\noindent
with $G^{0}(\omega)$ given by (\ref{G0}). When $a\gg1$
the expression (\ref{G-weak}) could be simplified at the above
condition $a \sin\Omega\gtrsim 1$. In this case we use (\ref{G0-asymp})
and observe that, upon differentiating over $\phi$, the main
contribution of order of $a$ stems from the numerator $\exp(\pm i
a\Omega)$ of (\ref{G0-asymp}).  Then one uses the identity $
\exp\left[\frac za \frac{\partial}{\partial a } a\right] =
\frac1a\exp\left[z \frac{\partial}{\partial a } \right] a $ and finds

     \begin{eqnarray}
     G(\omega) &\simeq& - \frac{ m_e}2
     \frac{\exp[ip\beta(\pi-\Omega)]}
     {\sqrt{2\pi a\sin\Omega}}
     \left[
     \frac{e^{ia(\pi-\Omega)-i\pi/4}}{\cos \pi(a+p\beta)}
     \right. \nonumber \\ &&\left.
     + \frac{e^{-ia(\pi-\Omega)+i\pi/4}}{\cos \pi(a-p\beta)}
     \right]
     \label{G-weak2}
     \end{eqnarray}

\noindent
with $\beta = \sin\theta\sin\theta'\sin\phi/\sin\Omega$. As before, in
the presence of damping $r_0> l_{mfp}$, the coherence on the sphere
breaks and the only surviving term in (\ref{G-weak2}) has the form

     \begin{equation}
     G_{damped}(\omega) \simeq
     \exp( {i \widehat{\bf z}({\bf r}'\times{\bf r})/2l_\ast^2} )
         G^{0}_{damped}(\omega),
     \label{G-weak2-d}
     \end{equation}
in accordance with previous results (see e.g.\ \cite{Kawabata}).

\section{the strong field regime}

Let us now turn to the case of strong fields, $p\to \infty$.
We define the integer number $n\geq0$ as follows : $2n = l-|m|$
for even $l-|m|$ and $2n+1=l-|m|$ for odd $l-|m|$. The value of $n$ has
a simple meaning, it corresponds to the number of zeroes of wave
function $S_{lm}(p,\cos\theta)$ within the interval $\theta \in
(0,\pi/2)$, by analogy with the known property of $P_l^m(\cos\theta)$.
The spectrum of (\ref{sphdifeq}) is then given by a series
\cite{Flammer,Me-Sch,KPSS}

     \begin{eqnarray}
     \varepsilon_{lm}(p) &=&
     4p \left[n+( m+|m|+1)/2 \right]
     - (s^2-m^2+1)/2
     \nonumber \\ &&
     +  O\left[s^3/ p \right],
     \label{ene-strong}
     \end{eqnarray}

\noindent
with $s=2n +|m| +1$ ; one has $s = l+1$ ($s=l$) for even (odd)
values of $l-m$. The eigenfunctions are given by

     \begin{eqnarray}
     S^\pm_{lm}(p,\eta) &=&{\widetilde S}_{lm}(p,\eta) \pm
     {\widetilde S}_{lm}(p,-\eta)  ,
     \label{wf-strong0}
     \end{eqnarray}
where plus (minus) sign corresponds to even (odd) values of $l-m$.
\cite{KPSS} The functions ${\widetilde S}_{lm}(p,\eta)$ are found as a
series in the Laguerre polynomials $L_n^m(x)$ ; in the main order of
$p^{-1}$ they can be written as

     \begin{eqnarray}
     {\widetilde S}_{lm}(p,\eta)
     &\simeq&
     c \,
     (1-\eta)^{|m|/2}
     e^{-p(1-\eta)}  L_n^{|m|}(2p(1-\eta)) ,
     \label{wf-strong1}
     \\
     c &=&
     \left[ \frac{2^{|m|}p^{|m|+1} n! }{2\pi r_0^2(n+|m|)!}
     \right]^{1/2}
     \nonumber
     \end{eqnarray}

\noindent
We see that in the main order of $p$ the Landau quantization takes
place, i.e.\ the spectrum (\ref{ene-strong}) is that of quantum
oscillator with the cyclotron frequency being the energy quantum.
\cite{LL3} The convergence and hence the applicability of the series
(\ref{ene-strong}) is given by the condition $p \gtrsim s\sim l$. In
its turn, it means \cite{Ba-Er} that all $n$ zeroes of the approximate
eigenfunction $L_n^{|m|}(2p(1-\eta))$ in (\ref{wf-strong1}) lie in the
northern hemisphere, $\eta>0$, as they should.

We notice the following important property of the spectrum
(\ref{ene-strong}). For given non-positive $m$ the values of
$\varepsilon_{lm}$ corresponding to $l=2n+|m|$ and $l=2n+|m|+1$
coincide. This property and the form of the wave function
(\ref{wf-strong0}) can be understood as follows. At $p\to
\infty $ the field-induced potential $p^2\sin^2\theta$  in
(\ref{sphdifeq}) localizes the particles near the poles of the sphere.
This form of two-well potential leads to the discussed degeneracy of
energy levels, while the total wave function is given by a
symmetrization (\ref{wf-strong0}) of the wave functions
(\ref{wf-strong1}) related to each of the wells. The possibility of
quantum tunneling between the wells lifts the degeneracy and produces
the exponentially small energy splitting ($\sim e^{-2p}$) between the
states $S^+_{lm}(p,\eta)$ and $S^-_{lm}(p,\eta)$ (\ref{wf-strong0}).
\cite{KPSS}

It would be tempting to obtain the Green's function at $p\to \infty$ in
the closed form. It can be done at the following simplifications.
First, we neglect the exponentially small splitting and notice that
$S^+(\eta) S^+(\eta') + S^-(\eta) S^-(\eta') = 2 [
{\widetilde S}(\eta) {\widetilde S}(\eta') +
{\widetilde S}(-\eta) {\widetilde S}(-\eta') ]$, i.e. the correlations
within one hemisphere only survive. Now we leave only the leading
terms (\ref{wf-strong1}) of the wave functions and $O(p)$ terms in
energies (\ref{ene-strong}). The necessity to restrict the summation
in (\ref{defG}) by the terms with $s\lesssim p$ could be modelled by
inclusion of the cutoff factor $e^{-s\delta}$ with $\delta \sim p^{-1}$
into (\ref{defG}). We put $\mu=0$ for simplicity of writing and raise
the denominator into the exponent, $(\omega  -E_{lm})^{-1} = i
\int_0^\infty dt\, e^{it(E_{lm}-\omega)} $.  Next we perform the
summation over $n$ with the use of bilinear generating function for the
Laguerre polynomials \cite{Ba-Er} :

     \begin{eqnarray}
     &&\sum_{n=0}^\infty
     \frac{n!}{(n+|m|)!} L_n^{|m|}(x) L_n^{|m|}(y) z^n
      \nonumber \\
     &&=  \frac{(xyz)^{-|m|/2}}{1-z}
     \exp\left( -z\frac{x+y}{1-z} \right)
     I_{|m|}\left( 2\frac{\sqrt{xyz}}{1-z} \right) .
     \nonumber
     \end{eqnarray}

\noindent
with the modified Bessel function $I_m(w)$; in our case $x=2p(1-\eta)$,
$y=2p(1-\eta')$ and $z=e^{it\omega_c}$. Making use of the property
$I_{|m|}(iw) = i^m J_m(w)$ and rescaling $t\omega_c/2 \to t$ we arrive
at the intermediate formula for $\eta,\eta' >0$ :

     \begin{eqnarray}
     G(\omega) &\simeq&
     - \frac{m_e}{4\pi}
     \int\limits_{0}^{\infty} \frac{dt }{\sin(t+i\delta) }
     e^{-2it\omega/\omega_c -\frac i2 (x+y) \cot(t+i\delta) }
     \nonumber \\ && \times
     \sum_{m=-\infty}^\infty
     e^{i(t-\phi+\pi/2)m} J_m\left( \frac{\sqrt{xy}}
     {\sin(t+i\delta) }\right)
     \nonumber
     \end{eqnarray}

\noindent
The summation over $m$ is now easily done ($\sum_m e^{im\phi} J_m(w)
= e^{iw\sin\phi}$) and we obtain the Green's function as a sum of two
terms, referring to two hemispheres. The ``northern'' term,
corresponding to the above expression, is given by

     \begin{eqnarray}
     G^{n}(\omega) &\simeq&
     - \frac{m_e}{4\pi}e^{iv}
     \int\limits_{i\delta}^{i\delta+\infty} \frac{dt }{\sin t}
     e^{-2i(t-i\delta)\omega/\omega_c -\frac i2\rho \cot t }
     \label{G-strong}      \\ &=&
     - \frac{m_e e^{iv -2\delta \omega/\omega_c} }
     {8\pi \cos \pi \frac{\omega}{\omega_c}}
     \int\limits_{-\pi/2 + i\delta}^{\pi/2+ i\delta}
     dt\,\frac{ \exp[-\frac{2i\omega}{\omega_c}t
     +\frac {i\rho\tan t}{2} ]
     }{ \cos t}
     \nonumber
     \end{eqnarray}

\noindent
with the analogue of the distance on the sphere $\rho = 2p[
2-\eta-\eta' - 2\sqrt{(1-\eta)(1-\eta')}\cos(\phi+i\delta)] $ and that
of the vector product $ v = 2p\sqrt{ (1-\eta) (1-\eta')}
\sin(\phi+i\delta) $. The form of the ``southern'' term is obtained by
changing $\eta\to-\eta$ and $\eta'\to-\eta'$ in these expressions.
The last integral is reduced for vanishing $\delta$ to the
hypergeometric function $\Psi(a,b;z)$ and we obtain the ``northern''
term in the form~ :

     \begin{eqnarray}
     G^{n}(\omega) &\simeq&
     - \frac{m_e}{4\pi}e^{iv-\rho/2
     -\delta \omega/\omega_c}
     \Gamma\left[\frac12 -\frac\omega{\omega_c}\right]
     \Psi\left[\frac12 -\frac\omega{\omega_c},1 ;\rho \right],
     \nonumber \\ &=&
     \frac{m_e}{4\pi}
      e^{iv-\rho/2 -\delta \omega/\omega_c}
     \sum_{n=0}^ {\infty}\frac{L_n(\rho)}{\omega/\omega_c-n-1/2}.
     \label{G-strong1}
     \end{eqnarray}
The last equation is similar to previous findings \cite{Kawabata}.

Let us discuss the applicability of (\ref{G-strong1}). Our derivation
was straightforward until (\ref{G-strong}), while the last step
demanded $\delta \sim 1/p \to 0$. The incomplete restructuring
of the spectrum into the Landau level scheme is absent in the usual
planar geometry, wherein one would put $\delta=0$ and the expression
(\ref{G-strong1}) would be exact. In the spherical case we cannot
treat the higher levels with $l \gtrsim p$ in an analytical way, it is
mimicked in (\ref{G-strong1}) by the appearance of the exponential
cutoff at $\omega \gtrsim p\omega_c$.

A subtler issue in justifying (\ref{G-strong1}) is the shift of the
last integration in (\ref{G-strong}) to the interval $(-\pi/2,\pi/2)$.
The integration over the remaining segments $(\pm\pi/2, \pm\pi/2
+i\delta)$ yields the factor $\cos(\pi \omega/\omega_c)$, thus the
contribution of these segments to the Green's function has no poles in
$\omega$. This contribution could be combined with the smooth (real)
part of $G(\omega)$ stemming from the consideration of the highest
energy levels.

As a result, one may conclude that the expression (\ref{G-strong1})
correctly reproduces the basic properties of the Green's function for
$\omega < p\omega_c$.

The finite value of the cutoff parameter $\delta$ in (\ref{G-strong1})
becomes important for the one-point correlation function, i.e. at
$\eta=\eta'$ and $\phi=0$. The residues of the Green's function define
the local density of states (LDOS) by the relation $N({\bf r}) = \int
d\omega\, n_F(\omega)[ G({\bf r},{\bf r},\omega-i0) - G({\bf r},{\bf
r},\omega+i0) ] /(2\pi i) $. If we assume that $n$ lowest Landau levels
are filled by the electrons, then it follows from (\ref{G-strong1})
that the LDOS is given by $N({\bf r})= n p e^{iv-\rho/2}/(2\pi r_0^2)$.
In this case the finite $\delta\sim 1/p$ provides a smooth variation of
LDOS of the form \[ N({\bf r}) \propto \exp[- O(\sin^2 \theta)], \]
which variation is absent in the usual planar geometry. This result,
however, may depend on the approximations made and requires further
numerical investigation.

\begin{figure}
\centerline{\epsfxsize=8cm
\epsfbox{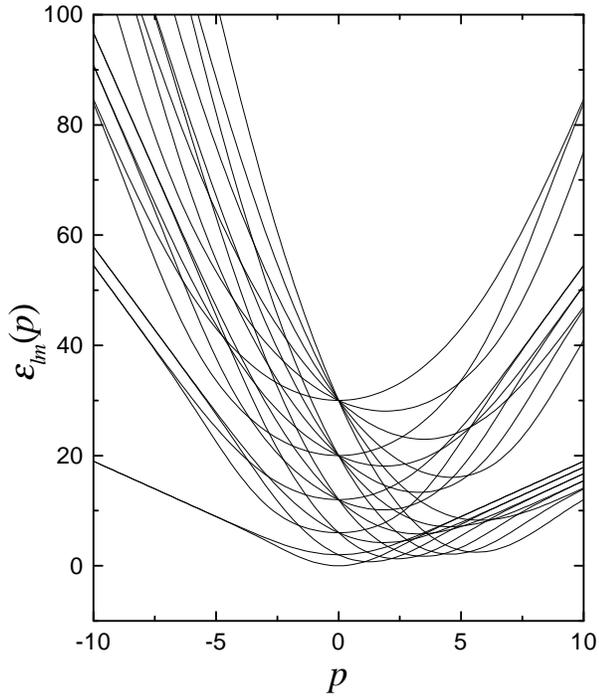}}
\caption{ The dependence of the energy levels $\varepsilon_{lm}$
on the number $p$ of magnetic flux quanta piercing the sphere.
For convenience of presentation, we plotted the evolution of
$\varepsilon_{lm}(p)$ with $m\leq 0$ on the rhs of the plot, at $p>0$.
The evolution of $\varepsilon_{lm}(p)$ with $m\geq 0$ is depicted on
the lhs of this plot for the formally negative $p$, see the text for
additional explanations.
\label{fig:lines}  } \end{figure}

It should be stressed that at the intermediate fields, at $p>1$ and
$p\lesssim l \lesssim p^2$, the spectrum and wave functions are not
principally reduced to closed expressions of hypergeometric type.
The numerical methods are indispensable here.
We calculated the evolution with $p$ of the energy levels with
$l=0,\ldots 5$ by diagonalizing the $200\times200$ tridiagonal matrices
in the basis of $P_l^m(\eta)$. The results are shown in the Fig.\
\ref{fig:lines} where
we made a following convention.
From the general property of Eq.(\ref{sphdifeq}) it follows that
$\varepsilon_{l,-m}(p) = \varepsilon_{l,m}(-p)$ with formally negative
$p$. \cite{fnote} For better readability of the graph, it is possible
then to show all the data, plotting only half of them with one sign of
$m$ but in the formally extended region of $p$.
 One can see in the
Fig.\ \ref{fig:lines} that
the degeneracy at $p=0$ is eventually changed by
the Landau levels formation at $p=10$. In the intermediate region
$p\sim 3$ the absence of any structure in the levels' scheme is noted.
The mesh of lines at the intermediate fields mimicks the chaotic
behavior, although this is not so. Both in the quantum problem
considered here and in its classical couterpart one has two variables
$(\theta,\phi) $ and two integrals of motion, the energy and
the projection of the angular momentum onto the field direction. To
obtain chaos, it suffices to break the rotation symmetry, $\phi\to \phi
+\delta\phi$. The latter problem is, however, beyond the scope of this
study.

In conclusion, we demonstrate the exact solution of the electron gas on
the sphere in the magnetic field. In the limits of weak and strong
fields this solution is reduced to the hypergeometric functions and the
observable quantities are found in the closed form. In the case of
intermediate fields, the solution is not essentially hypergeometric
function and the observables require further numerical analysis.

\acknowledgements

The author thanks S.V. Maleyev, M.L.Titov, S.L. Ginzburg, V.E. Bunakov,
V.I. Savichev for helpful discussions. This work was supported in part
by the RFBR Grant No.\ 96-02-18037-a and Russian State Program for
Statistical Physics (Grant VIII-2).


\end{multicols}

\vskip.4cm

\vskip.5cm


\begin{references}
\bibitem{fnote} This property is seen in the low field limit,
Eq.(\ref{ene-weak}), but is obscure in the expression for the
strong fields, Eq.(\ref{ene-strong}). The reason is that
the function $\varepsilon_{lm}(p)$ has cuts in the complex plane of
$p$, hence the analytical continuation $p\to -p$ is not straightforward
for $p$ near the complex infinity. \cite{Me-Sch}
\end{references}
\end{document}